\documentclass[11pt]{elsarticle}

\pdfoutput=1

\makeatletter
\def\ps@pprintTitle{%
  \let\@oddhead\@empty
  \let\@evenhead\@empty
  \let\@oddfoot\@empty
  \let\@evenfoot\@oddfoot
}
\makeatother

\usepackage{url}
\usepackage{breakurl}
\usepackage[breaklinks,
            colorlinks = true,
            linkcolor = blue,
            urlcolor  = blue,
            citecolor = blue,
            anchorcolor = blue]{hyperref}

\usepackage{lineno}

\usepackage{graphicx}
\usepackage[margin=1.25in]{geometry}
\usepackage[usenames,dvipsnames]{color}

\newcommand\acp{\begin{center}
\rule[-0.2in]{\hsize}{0.01in}\\\rule{\hsize}{0.01in}\\
\vskip 0.1in Submitted to the  Proceedings\\ 
of the African Conference on Fundamental and Applied Physics
    \vskip 0.05in
    {\it Second Edition, ACP2021, March 7--11, 2022 --- Virtual Event}\\
\rule{\hsize}{0.01in}\\\rule[+0.2in]{\hsize}{0.01in} \\
\end{center}}

\usepackage[firstpage=true]{background}
\backgroundsetup{contents={\parbox{6.5in}{\acp}}, scale=1,placement=top,opacity=1,color=black,position={3.25in,1.2in}}

\usepackage{fancyhdr}
\fancypagestyle{plain}{%
  \fancyhf{}%
  \fancyhead[C]{}
  \fancyfoot[C]{\thepage}
}

\fancypagestyle{empty}{%
  \fancyhf{}%
  \fancyhead[C]{{\it ACP2021, March 7--11, 2022 --- Virtual Edition}}
  \fancyfoot[C]{\thepage}
}
\begin{document}

\begin{frontmatter}


\title{High Energy Physics activities in Africa: An overview}

\author[add1]{Yasmine Amhis\corref{cor1}}
\ead{yasmine.amhis@ijclab.in2p3.fr}
\author[add2]{Mohamed Chabab\corref{cor2}}
\ead{mchabab@uca.ac.ma}
\author[add3]{Zinhle Buthelzi}

\cortext[cor2]{Corresponding Author}

\address[add1]{IJCLab, Orsay, France}
\address[add2]{ LPHEA, Cadi Ayyad University, Marrakech-Morocco }
\address[add3]{iThemba LABS, Soth Africa}

\begin{abstract}
\noindent 
    This document summarises our best knowledge of the ongoing High Energy Physics activities in Africa. The information was primarily extracted from the first ASFAP Particle Physics day organised on November 2021 and on our working group talk presented at ACP 2021 conference on March 2022. 
\end{abstract}


\end{frontmatter}

\section{Introduction}
\label{sec:intro}
\noindent

High Energy Physics (HEP) reveals the profound connections underlying all observed phenomena, ranging from the smallest to the largest structures in our Universe. Everything in our universe is found to be made from elementary particles, as a few basic matter blocks, governed by four fundamental interactions.  Our best knowledge of how these particles interact is encoded in the Standard Model of particle physics (SM). The SM developed in the seventies has become an established and well tested theory. This document is divided in two section, the first one focuses on  theoretical HEP physics while  the second one is dedicated to  experimental particle physics in Africa. For each field, the activities are reported per country by alphabetical order. If an important activity had been forgotten please contact the authors. To contribute in a significant way to the development of HEP in Africa, we believe that we should focus on maintaining  leadership of the organization of HEP education programs in some targeted institutes, with involvement of African governments and policy makers.
To this end, ASFAP has dedicated a working group to  Particle Physics with the aim to build an African network,   support and expand the activities in this field, and ultimately prepare a road map based on collected letter of intents proposed by particle physics community. The list provided in Table~\ref{tab:activities} is a tentative summary of  the current (as of 2022) involvements of African countries in particle physics experiments.  

\section{Overview on Theoretical physics in Africa}
\label{sec:theory}
 In July 2012, ATLAS and CMS experiments at LHC have announced the discovery of a scalar particle, later identified as a Higgs boson, the last missing piece of the Standard Model \cite{atlas_2012, cms_2012}. However, despite its success, there are still many fundamental questions awaiting a clear answer, which require the construction of new theoretical models,  beyond the SM, which is then treated as  effective theory of a more fundamental description. Among  unsolved problems and experimental data that cannot be explained by the SM , we can cite: the neutrino mass generation, pattern of fermions hierarchy, and dark matter/energy.  
 possible extensions of the SM 
 This means that we have to look at alternatives beyond Standard Model (BSM) that could solve those issues and could be tested at the LHC and future colliders. So far, many theoretical proposals are on the shelves: The most attractive one are BSM Models involving Higgs fields with higher representation: more doublet, more singlet, doublet and triplet,..
 
 Several African groups have strong expertise in phenomenology of the beyond the Standard Model Physics. Theses groups have gained footholds in dealing with multi-Higgs models with an emphasis on the symmetries of the scalar sector and Dark Matter candidates arising from these models.
 
 The Egyptian team, from the Center for Fundamental Physics (CFP) at Zewail City Research areas, is specialist in many high energy theory topics including: Susy phenomenology, early Universe and astro-Particle Physics (selected references \cite{zuwail}). 
 
 The Moroccan groups, from Cadi Ayyad and Abdelmalek Saadi Universities, have relevant expertise in phenomenological studies in BSM physics including non-minimal Higgs models and supersymmetric scenarios with a particular focus on Higgs phyics , theoretical and EW precision constraints on scalar sectors in various extensions of the SM, including their implementation in high energy physics tools (selected references \cite{morocco}). 
 
 The East African Institute for Fundamental Research (EAIFR), at the University of Rwanda has research interest in fundamental physics with a focus on collider physics, physics beyond the Standard Model, cosmic inflation, Dark Matter and Dark Energy. EAIFR has produced significant papers on the impact of additional Higgs bosons on signal rates and study of possible deviations from the SM (selected references \cite{rwanda}). 

The South African HEP groups are strongly involved in development of BSM phenomenology and analysis of the data collected by the ATLAS experiment at the LHC. They are mainly affiliated to U. Witwaters, Johansburg U. and iThemba LABS (selected references \cite{sa}). 

At last, a team from Madagascar is specialist of non perturbative methods in strong interactions. More precisely, they use QCD sum rules to predict hadron properties, such as masses and coupling constants(selected references \cite{mad}).

\noindent

\section{Experimental physics}
\label{sec:experimental}
\begin{figure}
\includegraphics[width=0.48\textwidth]{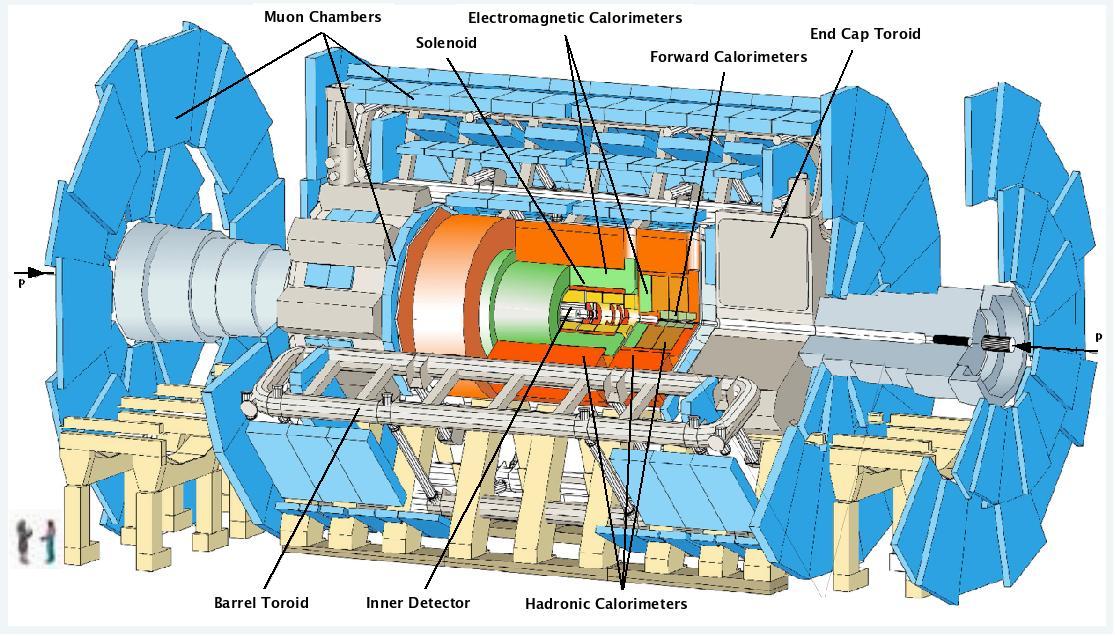}
\includegraphics[width =0.48\textwidth]{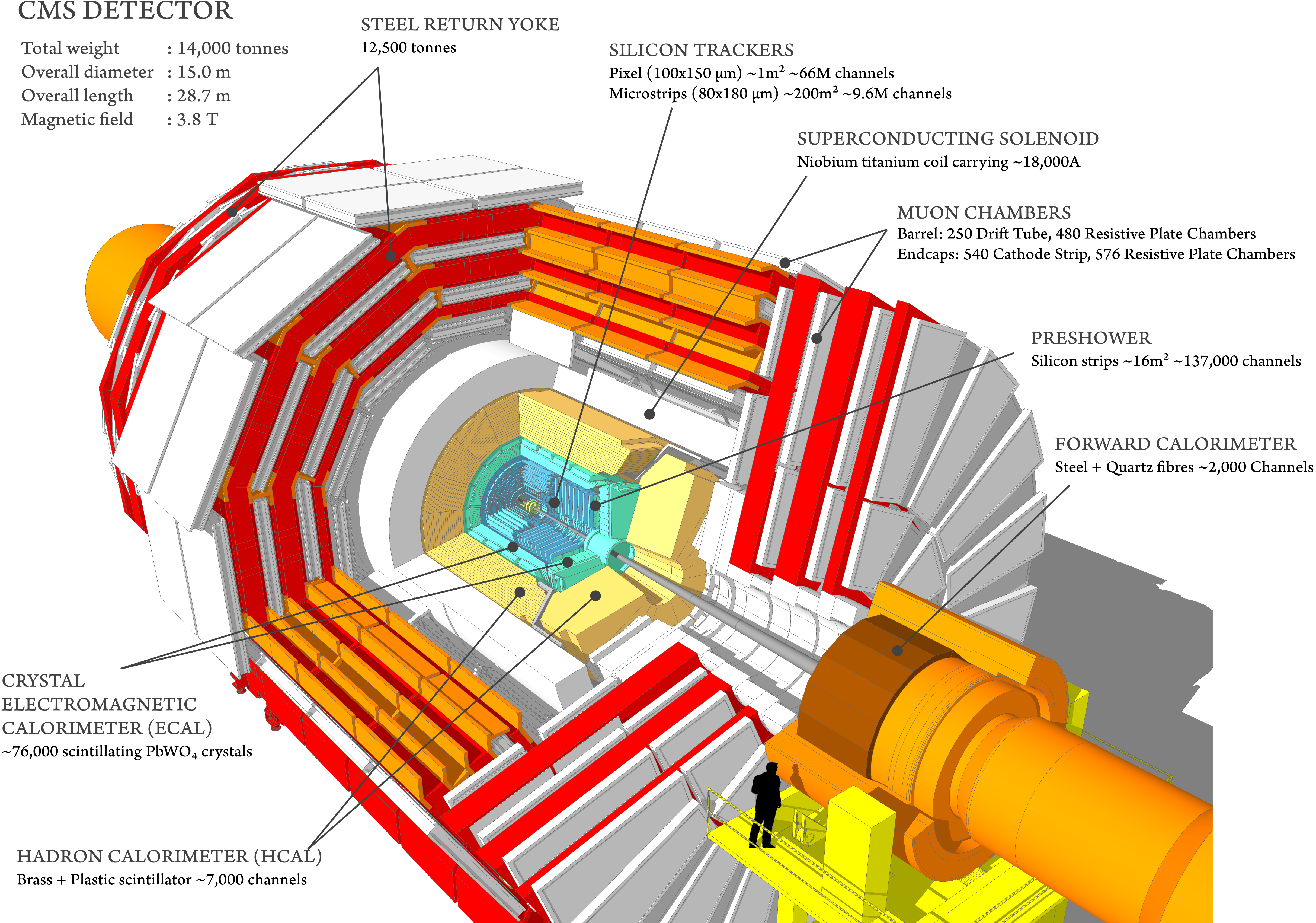}
\caption {ATLAS (left) and CMS(right) detectors.}
\label{fig:detectors}
\end{figure}

The Large Hadron Collider is the largest and most powerful collider in the world. It is located at CERN between Switzerland and France. The first proton beams started to circulate in 2008. Four major experiments, ATLAS, CMS, LHCb and ALICE are located across the ring. A sketch of ATLAS and CMS can be found in Figure ~\ref{fig:detectors}.
Their purpose is complementary and aims at understanding the behaviour of fundamental particles and their interactions. An upgrade of the LHC, HL-LHC, is foreseen in 2025.  After a successful period of data taking (Run1/2) the detectors are being upgraded in many phases. A detailed list of the associated institutes can be found in Table~\ref{tab:activities}.  
The information collected in this document is based on the material that was presented at the   \href{https://indico.cern.ch/event/1080353/timetable/#20211118.detailed}{First ASFAP Particle Physics Day}.

\label{sec:grate}
\noindent
\subsection{Algeria} 
A computing group contributes to ATLAS to face future computing challenges during the HL-LHC upgrade. This body of work consists of two projects porting of ATLAS software to parallel architectures and monitoring of conditions database access.

\subsection{Egypt}
The cluster of groups contributes to the CMS experiment.  On the physics analyses side, Beyond Standard Model searches have been or are being conducted. For instance a $Z^\prime$ search,  Kaluza Klien excitation from Extra-dimensions, and also Dark matter searches signatures combined with  mono-$Z$ mono-Higgs or mono-$Z^{\prime}$ topologies. 
On the detector side the groups are involved in developments of the Resistive Plate Chamber (RPC) as well as Gas Electron Multiplier (GEM). 

\subsection{Madagascar}

DUNE is an international flagship experiment to unlock the mysteries of neutrinos. The group contributed to the Near Detector Conceptual design report and to the SAND-System for on-Axis Neutrino Detection.

\subsection{Morocco}

\paragraph{ATLAS:}
In 1996, Morocco signed an agreement with CERN and became the first African member of the ATLAS collaboration.  The RUPHE,  a Moroccan cluster of several HEP groups, works currently on HL-LHC, High Granularity Timing Detector (HGTD). Two axes are pursued on the $b$-tagging and performance optimisation. For the first axis, there are ongoing studies are based on the Generic Boosted Decision Tree and
self-tagging.  For the latter the aim is improvement the jet energy resolution in the forward region and use of particle-flow reconstruction and HGTD impact.
The assembly of HGTD modedules locally coordinated by MAScIR. 
The groups contributed team beam campaigns to finalise the  choice of Low Gain Avalanche Detector (LGAD).  
On the analyses side the following topics have been investigated: 
Search for a BSM resonance in the top quark sector, search for Invisible Higgs, 
Search for Hidden Higgs or dark Sector, search for charged Higgs in $H^{+}\to tb$ channel, diboson resonances in semi-leptonic final states $Z^\prime$, Kaluza Klein and Dihiggs $HH$ resonant and non-resonant production : $bb\ell\ell$ and $bbVV$ channels

\paragraph{ANTARES/KM3Net}
KM3NeT, the legitimate successor of ANTARES, is a new research nfrastructure consisting of a network of deep-sea neutrino telescopes in the Mediterranean Sea. The main objectives of the KM3NeT1 Collaboration are: i) the discovery and subsequent observation of high- energy neutrino sources in the Universe and ii) the determination of the mass hierarchy of neutrinos (MHN). These objectives are strongly motivated by two recent important discoveries, namely: The high- energy astrophysical neutrino signal reported by IceCube,  and the sizeable contribution of electron neutrinos to the third neutrino mass eigenstate as reported by Daya Bay, Reno and others. To meet these objectives, KM3NeT is building two detectors ORCA and ARCA. Morocco has signed an agreement to join KM3NeT collaboration in 2017. So far three universities (Mohammed V U., Cadi Ayyad U., and Mohammed 1 U.),  currently full members of the collaboration, are actively  participating in the production line of optical modules in a national site located in Rabat.  Besides, the Moroccan team is also involved in the physics analysis of many topics, essentially related to search for magnetic monopoles, search for nuclearites, and study of the neutrino mass hierarchy \cite{km3net}.

\subsection{South Africa}
There are multiple South African experimental HEP research groups active in both the ALICE and ATLAS experiments.  

\paragraph{ALICE} The group contributes to upgrade projects towards a common read out unit for the muon identifier, the Low-Voltage System for muon tracking, and online data processing for the Transition Radiation Detector. Given the travel restrictions, the possibility to work operate the systems remotely has been utilised. The ALICE experiment explores the outcomes of heavy ion collision, the group worked on $W$ and $Z$ boson tests of the Standard Model via the study of the cross-sections in lead-lead and proton-lead collisions.

\paragraph{ATLAS} On the hardware side the following activities are ongoing:
\begin{itemize}
    \item Silicon detector developments on both the SCT and ITk system including, data acquisition electronics development, evaporative cooling systems, material description in simulation, firmware and test QC for EoS redout cards,  polymoderator design, procurement, and fabrication.
    \item Muon New Small Wheel work including, material description in simulation, manufacturing and assembly of components and installation tools as well as commissioning. 
    \item ATLAS Local Trigger Interface boards were installed in the TTC crates of LBA, LBC, EBA, EBC and the Laser crate.
    \item Assembly, quality checks and installation of the gap scintillator counters on the ATLAS detector
    \item Phase-II upgrade of the Tile Calorimeter, 50\% of the production of the Low Voltage Power Supplies (LVPS), 24\% of the production of the Tile Preprocessor (PPr).
    \item Participation to ATLAS TileCal November 2021 Test-beam.
    \item CFD simulations for temperature and humidity distributions inside the detector ITk volume.
    \item Operation of the TDAQ SysAdmin and Network, Muon ConfigDB in the Control Room
    \item Detector Lab – Micro-Megas NSW.
\end{itemize}
On the physics analyses side, the following analyses are or have been pursued:
\begin{itemize}
\item Top quark mass measurement utilising leptonic $J/\psi$ decays. 
\item Higgs boson production in association with a $W/Z$ boson, with the Higgs decaying to two bottom quarks.
\item New Physics searches via the study of top electro-weak couplings in rare processes (ttW, tWZ)
\item Boosted Heavy Neutrino Search.
\item Dark and semi-visible jets: unusual signatures emanating from strongly interacting dark sector.
\item Anatomy of the multi-lepton anomalies.
\item The Higgs Portal to the Dark and or Hidden sector for example $H \to  Z_d Z_d \to 4e, 4\mu ,2 e2\mu$, $H\to \gamma\gamma_d$
\end{itemize}

\begin{table}[h]
\centering 
\scriptsize
    \begin{tabular}{c|l|c}
          Experiment & Institution & Country  \\
         \hline
          ANTARES & Faculté des Sciences, Université Mohammed 1, Oujda & Morocco \\
          ANTARES & Faculté des Sciences, Université Mohammed V, Rabat & Morocco \\
          ANTARES & Faculté des Sciences Semlalia, Université Cadi Ayyad, Marrakech & Morocco  \\
             \hline
        KM3Net & Faculté des Sciences Semlalia, Université Cadi Ayyad, Marrakech & Morocco \\
        KM3NeT & Faculté des Sciences, Université Mohammed 1, Oujda & Morocco \\
        KM3Net & Faculté des Sciences, Université Mohammed V, Rabat & Morocco\\
          KM3Net & Universities of Johannesburg/Witwatersrand/North-West & South Africa \\
                  \hline
          DUNE & The University of Antananarivo  & Madagascar \\
        \hline
        ATLAS Morocco Cluster & Faculté des Sciences Ain Chock, Université Hassan II, Casablanca & Morocco \\ 
        & Faculté des Sciences, Université Ibn-Tofail, Kénitra  & Morocco\\
         & LPHEA, Faculté des Sciences Semlalia, Université Cadi Ayyad, Marrakech  & Morocco \\  
         & LPMR, Faculté des Sciences, Université Mohamed Premier, Oujda  & Morocco \\ 
         & Faculté des sciences, Université Mohammed V, Rabat  & Morocco \\
         & Mohammed VI Polytechnic University, Ben Guerir  & Morocco \\
         \hline
         ATLAS South Africa Cluster &  Department of Physics, University of Cape Town, Cape Town & South Africa \\
         & Department of Mechanical Engineering Science, University of Johannesburg & South Africa \\
         & University of South Africa, Department of Physics, Pretoria & South Africa \\
         & iThemba Labs, Western Cape & South Africa \\
& University of South Africa, Department of Physics, Pretoria & South Africa \\
& University of Zululand, KwaDlangezwa & South Africa \\ & School of Physics, University of the Witwatersrand, Johannesburg & South Africa \\
          \hline
          ATLAS Technical Associate Institute & Ecole Nationale Supérieure d’Informatique (ESI)& Algeria\\ 
         \hline
            CMS & Academy of Scientific Research and Technology, Cairo & Egypt \\
           CMS & Center for High Energy Physics (CHEP-FU), Fayoum University, El-Fayoum & Egypt \\
           \hline
      ALICE & iThemba LABS, Universities of Cape Town/Witwatersrand & South Africa \\
    \end{tabular}
    \caption{Overview of ongoing High Energy Physics activities and institutions in Africa.}
    \label{tab:activities}
\end{table}

\section*{Acknowledgments}
The authors would like to thank Abdeslam Hoummada, Peter Jenni and James Keaveney for the fruitful discussions and feedback.

\bibliography{myreferences} 
\bibliographystyle{elsarticle-num}

\end{document}